\documentclass[sensors,article,submit,moreauthors,dvipdfm,12pt,a4paper]{mdpi}
\usepackage{amssymb,amsmath}
\usepackage{epsfig}
\usepackage{hyperref}
\usepackage{color}

\newcommand{\highlight}{\color{black}}

\title{Material Limitations on the Detection Limit in Refractometry\footnote{Special Issue "Laser Spectroscopy and Sensing", Edited by Prof. M.W. Sigrist}}

\author{Peder Skafte-Pedersen,$^{1}$ Pedro S. Nunes,$^{1}$ Sanshui Xiao,$^{2}$ and Niels Asger Mortensen $^{2,\star}$}

\address{%
$^{1}$ Department of Micro and Nanotechnology, Technical University of Denmark, DTU Nanotech, Building 345 East, DK-2800 Kongens Lyngby, Denmark\\
$^{2}$ Department of Photonics Engineering, Technical University of Denmark, DTU Fotonik, Building 345 West, DK-2800 Kongens Lyngby, Denmark\\
}
\emails{asger@mailaps.org}
\corres{}%asger@mailaps.org}

\abstract{We discuss the detection limit for refractometric sensors relying on high-$Q$ optical cavities and show that the ultimate classical detection limit is given by
$\min\left\{\Delta n\right\}\gtrsim \eta$ with $n+i\eta$ being the complex refractive index of the material under refractometric investigation. {\highlight Taking finite $Q$ factors and filling fractions into account, the detection limit declines. As an example we discuss the fundamental limits of silicon-based high-$Q$ resonators, such as photonic crystal resonators, for sensing in a bio-liquid environment, such as a water buffer.} In the transparency window ($\lambda \gtrsim 1100$~nm) of silicon the detection limit becomes almost independent on the filling fraction, while in the visible, the detection limit depends strongly on the filling fraction because silicon absorbs strongly.}

\begin{document}
\section{Introduction}
Refractometry is one of the classical workhorses among a variety of optical techniques in analytical chemistry. In its basic principle it allows the quantification of concentration changes by means of an associated refractive-index change leading to frequency shifts of optical resonances. Optofluidics integration~\cite{Psaltis:2006,Monat:2007,Erickson:2008} holds promises for refractometric analysis of minute sample volumes by the aid of optical resonators integrated in microfluidic architectures. Integrated nanophotonic resonators are beginning to show promising potential for high sensitivity and detection of minute concentrations~\cite{Schmidt:2004,Erickson:2006,Skivesen:2007,Nunes:2008,Mandal:2008} and the sensing performance is under active consideration in the research community~\cite{White:2008,Fan:2008,Mortensen:2008}. More recent photonic crystal resonator designs emphasize the optimization of the light-matter overlap~\cite{Kwon:2008} which serves to increase the sensitivity~\cite{Mortensen:2008}. However, the detection limit is in many {\highlight applications} of equal importance~\cite{White:2008}. \emph{How small a refractive-index change can a resonator-based sensor setup quantify reliably}? {\highlight In this paper we work out the material absorption limitations. This allows us to estimate the ultimate detection limit, provided that resonators with sufficiently high intrinsic $Q_0$ value are available.} For silicon based sensors operating in a water environment we find that for $\lambda \gtrsim 1100\,{\rm nm}$, the ultimate detection limit is given by $\min\left\{\Delta n\right\}\sim\eta$ with $\eta$ being the strongly wavelength-dependent imaginary index of water, i.e. the extinction coefficient. {\highlight As a consequence, it should in principle be possible to detect refractive-index changes down to $10^{-5}$ at $\lambda\sim 1200\,{\rm nm}$, while at wavelength of $\lambda\sim 1500\,{\rm nm}$ the ultimate detection limit is increased by an order of magnitude to $10^{-4}$. On the other hand, in the visible it is difficult to go below $10^{-2}$.} {\highlight According to our knowledge, the limitation of material absorption is a central, but overlooked issue that was only pointed out very recently in independent work (on photonic crystal resonators) by Tomljenovic-Hanic \emph{et al.}~\cite{Tomljenovic-Hanic:2009}.}

The remaining part of the manuscript is organized as follows. In Section~\ref{sec:theory} we use electromagnetic perturbation theory to calculate the sensitivity and the associated detection limit. In Section~\ref{sec:examples} we discuss our results in the context of various resonator examples and as a particular example we consider the ultimate detection limit for silicon-based sensors in a water environment. Finally, in Section~\ref{sec:conclusion} discussions and conclusions are given.

\section{Theory}
\label{sec:theory}

Consider an electromagnetic resonance with a density of states (or power spectrum) which for simplicity could be given by a Lorentzian line shape

\begin{equation}\label{eq:resonance}
\rho(\omega)= \frac{1}{\pi} \frac{\delta\omega/2}{(\omega-\Omega)^2+(\delta\omega/2)^2}
\end{equation}
where $\Omega$ is the resonance frequency and $\delta\omega$ is the line width corresponding to a quality factor $Q=\Omega/\delta\omega$. The sensitivity of a resonator is a measure of the resonance wavelength shift as function of the refractive-index change. For applications in refractometry, first order perturbation theory is adequate and gives~(see e.g. Reference~\cite{Mortensen:2008})
\begin{equation}\label{eq:perturbationtheory}
\Delta\Omega=-\frac{\Omega}{2}\frac{\big< E\big|\Delta \varepsilon\big|E\big>}{\big< E\big|\varepsilon\big|E\big>}.
\end{equation}
{\highlight This expression} can be used to calculate the resonance frequency shift caused by a small change in the real part of the complex refractive index for materials in proximity with the cavity mode. We label the different material constituents by the index $j$ so that~\cite{Mortensen:2008}
\begin{equation}\label{eq:sensitivity}
\Delta\Omega=-\Omega \sum_j f_j\frac{\Delta n_j}{n_j}
\end{equation}
where $n_j$ is the real part of the complex refractive index $n_j+i\eta_j$ and the filling fraction is given by
\begin{equation}
f_j= \frac{\big< E\big|\varepsilon\big|E\big>_j}{\big< E\big|\varepsilon\big|E\big>}
\end{equation}
with $\sum_j f_j=1$. The subscript in the numerator indicates that the integral is restricted to the volume fraction where the perturbation is present, while the integral in the denominator is unrestricted.

Next, consider refractometry where a small change in the real part of the refractive index in, say, material $j=1$ causes a shift in the resonance frequency. One first important question is of course what the sensitivity (or the responsivity) of the system is. The answer is given by Equation~(\ref{eq:sensitivity}) and basically the higher a value of $f_1$ the higher the sensitivity. However, in many applications, the detection limit is of equal concern. \emph{How small changes may one quantify}? {\highlight As discussed in Refs.~\cite{Tomljenovic-Hanic:2009,Armani:2007}, the resonance line-width $\delta\omega=\Omega/Q$ represents an ultimate measure of the smallest frequency shift that can be quantified accurately. Eq.~(\ref{eq:sensitivity}) consequently leads to a bound on the smallest refractive-index change that can be quantified accurately. In this way we arrive at}
\begin{equation}\label{eq:DnQ}
\min\left\{\Delta n_j\right\}\gtrsim \frac{n_j}{2f_jQ}.
\end{equation}
Obviously, the higher a quality factor the lower a detection limit. In the following we consider the general situation with
\begin{equation}
Q^{-1}=Q_0^{-1}+Q_{\rm abs}^{-1}
\end{equation}
where the first term corresponds to the intrinsic quality factor in the absence of absorption while the second term accounts for material absorption. For weak absorption ($\eta\ll n$) we apply Equation~(\ref{eq:perturbationtheory}), so that $\eta$ gives a small imaginary frequency shift. In the framework of Equation~(\ref{eq:resonance}) this causes an additional broadening corresponding to~\cite{Mortensen:2008,Xu:2009}
\begin{equation}
Q_{\rm abs}^{-1}=\sum_j  2f_j \frac{\eta_j}{n_j}.
\end{equation}
{\highlight The detection limit, Equation~(\ref{eq:DnQ}), now becomes}
\begin{equation}
\min\left\{\Delta n_j\right\}\gtrsim \frac{n_j}{2f_jQ_0} + \sum_i  \frac{f_i}{f_j} \frac{n_j}{n_i} \eta_i.
\end{equation}
This is the main {\highlight result of this section.} For a two-component structure with $f_1+f_2=1$, the result further simplifies to
\begin{equation}\label{eq:ultimate}
\min\left\{\Delta n_1\right\}\gtrsim \frac{n_1}{2fQ_0} + \eta_1 + \frac{1-f}{f} \frac{n_2}{n_1} \eta_2
\end{equation}
where we have introduced $f\equiv f_1$ and where the perturbation is assumed to be applied to medium $j=1$. We emphasize that a much similar result was reported recently by Tomljenovic-Hanic \emph{et al.}~\cite{Tomljenovic-Hanic:2009} in a study of silicon photonic crystal resonators, though $\min\left\{\Delta n_j\right\}$ was emphasized less explicitly. In this work we arrive at the simple result for $\min\left\{\Delta n_j\right\}$ by using the conventional limit, that the smallest detectable frequency shift is limited by the resonance linewidth, as also discussed by Reference~\cite{Tomljenovic-Hanic:2009}. If the signal-to-noise ratio is adequate, detection of sub-linewidth shifts is in principle possible, but this would call for more advanced data analysis, e.g. locking onto the resonant frequency in order to compensate for noise and fluctuations in the spectrum. As pointed out in Reference~\cite{Armani:2007}, the above classical limit is the most desirable one to consider when balancing detection limits and experimental complexity. {\highlight We note that our assumption of a homogeneous sample greatly simplifies the calculation while a general expression can not be derived for an inhomogeneous sample. In that case, one would need detailed information on the spatial variations of both the sample concentration as well as the field profiles.} In the following we analyze the consequences of Equation~(\ref{eq:ultimate}) for a number of optical sensing architectures as well as in the context of material parameters.

\section{Examples}
\label{sec:examples}

\subsection{Gas sensing}

Gas sensing and detection is an important application, which is however made difficult for various reasons including the fact that a dilute gas only perturbs the resonator modestly compared to vacuum conditions. This is very pronounced in the case where the resonator is detuned from strong gas-absorption lines so that $n\simeq 1$ and $\eta\simeq 0$. Glass-sphere resonators are in this context particularly interesting. They support high-$Q$ whispering-gallery modes which may be used for sensing~\cite{Armani:2007} and in particular, such resonators may be employed for refractometric measurements in a gaseous environments without strongly degrading the quality factor. {\highlight The cavity field is mainly confined to the interior of the sphere while the refractive-index perturbation occurs in the exterior, where the field is evanescent. This means that $f$ is vanishing and therefore the sensitivity is not particularly high, see Equation~(\ref{eq:sensitivity}).} However, in applications involving the detection of minute concentration changes this may be compensated by the possibilities for extremely low detection limits. In the limit with strong light confinement, Equation~(\ref{eq:ultimate}) simplifies to
\begin{equation}
\min\left\{\Delta n\right\}\gtrsim \frac{n}{fQ_0} + \eta ,\quad(f\ll 1).
\end{equation}
If measurements are performed away from a strong gas-absorption line, then $n\simeq 1$ and $\eta\simeq 0$ and the detection limit is basically limited by $1/fQ_0$ of the spherical resonator. On the other hand, if $\eta \gg 1/fQ_0$ then the detection limit is governed by the absorption being intrinsic to the gas. Photonic crystal structures may support resonances with values of $fQ_0$ in excess of $10^{5}$~\cite{Tomljenovic-Hanic:2009} and spherical resonators may potentially support even higher values~\cite{Arnold:2003,Teraoka:2003,Hanumegowda:2005,Armani:2007}. This makes refractometric measurements an important alternative to Beer--Lambert gas absorption measurements (see e.g. Reference~\cite{Jensen:2008} and references therein).

\begin{figure}[t]
\begin{center}
\epsfig{file=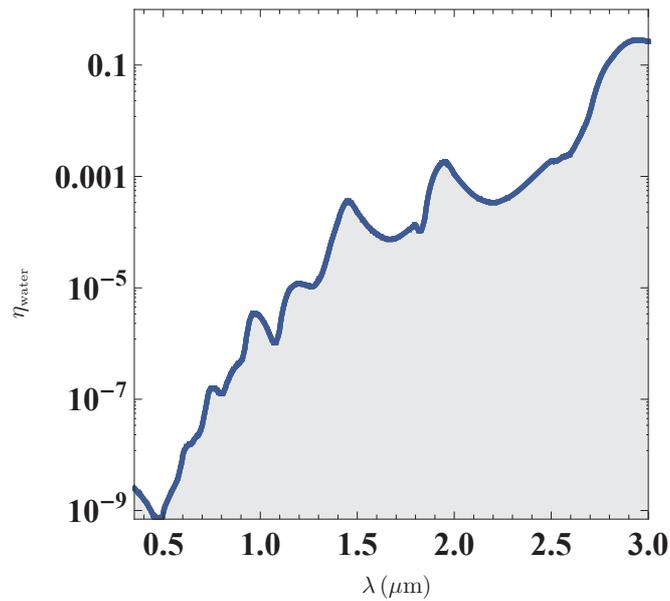, width=0.5\columnwidth,clip}
\end{center}
\caption{Imaginary part of the refractive index for water. According to Equation~(\ref{eq:water}), the curve also represents the ultimate detection limit for refractometry in a water environment. Data is reproduced from Reference~\cite{Querry:1991}.} \label{fig1}
\end{figure}

\subsection{Liquid-phase sensing --- general considerations}

Imagine that sensing occurs in the liquid phase, so that the damping of the liquid cannot be ignored. In the following we consider water which is the natural environment and buffer for a range of bio-chemical applications. From our general analysis it follows that one can never achieve a detection limit that overcomes the damping in the liquid. {\highlight In the limit of perfect light-liquid overlap ($f\rightarrow 1$) and a high intrinsic quality factor ($Q_0\gg 1/\eta_{\rm water}$), Equation~(\ref{eq:sensitivity}) gives}
\begin{equation}\label{eq:water}
\min\left\{\Delta n\right\} > \eta_{\scriptscriptstyle \rm water}.
\end{equation}
In Figure~\ref{fig1} we illustrate the implications of this result by showing experimental data for water~\cite{Querry:1991}. Experimentally, $\eta$ may be obtained from the exponential damping in a transmission measurement where the absorption coefficient is given by $\alpha_{\rm abs}=\eta 4\pi/\lambda$. We emphasize that for any resonator design, the detection limit will tend to be higher unless the resonance line-width is dominated by liquid absorption. Below we offer a few examples of resonator architectures which have been considered in the literature.

\subsection{Liquid-droplet resonators}

{\highlight Consider a liquid-droplet resonator where the droplet radius $R$ is sufficiently small that gravity can be neglected. Due to surface tension, the droplet now forms a perfectly spherical resonator with the total-internal reflection mechanism supporting whispering gallery modes inside the liquid volume.} Because of the total-internal reflection the cavity field is confined to the same volume as where the refractive-index perturbation occurs and thus the resonances have $f\simeq 1$. The above analysis can now be used to estimate the ultimate detection limit for refractive-index changes in the liquid,
\begin{equation}
\min\left\{\Delta n\right\}\gtrsim Q^{-1} \simeq \eta_{\scriptscriptstyle\rm water}  \quad (Q_0\gg Q_{\rm abs}).
\end{equation}
The data in Figure~\ref{fig1} for water makes such liquid-droplet particularly interesting for sensing applications in the visible part of the spectrum.
Liquid droplet are considered interesting candidates for high-Q cavities, because the whispering-gallery modes do not suffer from surface roughness, as surface-tension makes the surface shape spherical and smooth on the molecular length scale. The present analysis suggests that these systems my potentially provide the ultimate detection limit for liquid-phase refractometry. Liquid-droplet resonators have recently been the subject of a number of studies~\cite{Azzouz:2006,HosseinZadeh:2006,Kiraz:2007,Kiraz:2007a} and for a recent review of droplet-based cavities and resonators, we refer to Reference~\cite{Molhave:2009}.

\begin{figure}[t]
\begin{center}
\epsfig{file=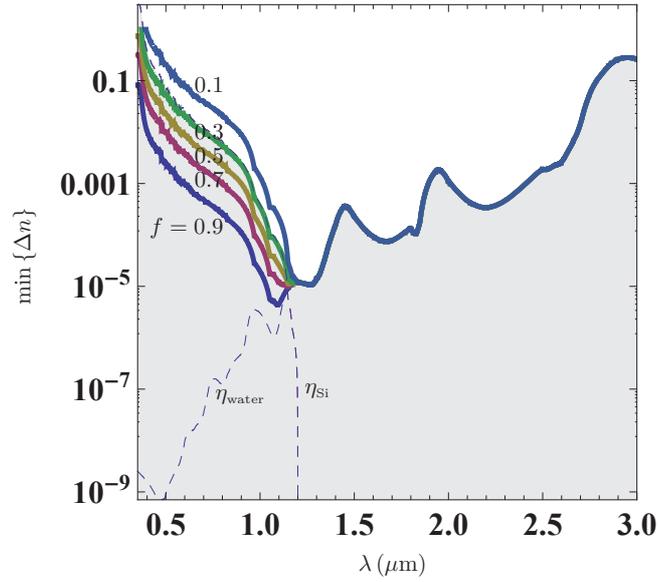, width=0.5\columnwidth,clip}
\caption{Ultimate detection limit for a resonance in a silicon structure infiltrated by water. The different traces show results for varying values of the light-liquid overlap $f$. The dashes lines shows the imaginary part of the refractive index for water and silicon, respectively. Based on data for water and silicon from Refs.~\cite{Edwards:1985,Querry:1991}.} \label{fig2}
\end{center}
\end{figure}

\subsection{Liquid-infiltrated silicon photonic crystals}

As an example we consider silicon photonic crystals (Si) infiltrated by liquid with a complex refractive index. Assuming that the intrinsic line width can be neglected we get

\begin{equation}
\min\left\{\Delta n\right\}\gtrsim \eta_{\scriptscriptstyle \rm water} +   \frac{1-f}{f} \frac{n}{n_{\scriptscriptstyle \rm Si}} \eta_{\scriptscriptstyle \rm Si} \quad (Q_0\gg Q_{\rm abs}).
\end{equation}

Since the imaginary part of the refractive index is strongly dispersive for both water and silicon, the ultimate detection limit clearly depends on the operational wavelength. Furthermore, the fact that water is most transparent in the visible, while silicon transmits best in the near-infrared, suggests that there exists an optimal wavelength for the detection limit in terms of the refractive-index change. Figure~\ref{fig2} illustrates the ultimate detection limit for silicon-based resonator structures infiltrated by liquid for varying values of the light-liquid overlap $f$. Calculations are based on data for water and silicon tabulated in the handbooks by Palik~\cite{Querry:1991,Edwards:1985}. Quite interestingly, the plot reveals that an optimal detection limit of the order $10^{-5}$ occurring around the onset of transparency in silicon at $\lambda\sim 1.1~{\rm \mu m}$. We emphasize that this ultimate detection limit of course implicitly assumes that the resonance has $Q_0> 10^5$. So far, quality factors of the order $Q\sim 5 \times 10^4$ have been reported in microfluidic silicon-based photonic crystal resonators~\cite{Bog:2008}, though the liquid was not water. It is also interesting to note that despite the transparency of water in the visible, the damping in silicon jeopardizes any realistic application with $f\lesssim0.9$.

Of course, if the detection limit is associated with the concentration $\cal C$ of molecules instead, one should also take the wavelength dependence of the so-called molar extinction coefficient [$\propto \lambda^{-1} \partial(\Delta n)/\partial{\cal C}$] into account, i.e. at which wavelength does a concentration change lead to the largest possible refractive-index change. Clearly, the extinction coefficient is specific to the molecule in question.

\section{Conclusion}
\label{sec:conclusion}

In conclusion, we have with the aid of perturbation theory worked out the fundamental limitations on the detection limit due to material absorption. As a general result, it is difficult to detect refractive-index changes smaller than the associated imaginary part of the refractive index. The main assumption behind this quite intuitive result is that the smallest detectable frequency shift is limited by the resonance linewidth. For liquid-infiltrated silicon-based optical resonators in e.g. photonic crystal architectures, we find an optimal wavelength range resulting from the strong wavelength dependence of $\eta$ for both water and silicon. Our results may be extended to also other material platforms. As an example, for visible applications one may take advantage of transparent glasses, such as SiON or SiO$_2$ where $\eta$ can be of the order of $10^{-4}$, though with a real part of the refractive index which is much reduced as compared to semiconductor materials like silicon. In the quest for refractometric sensors with yet better detection limits, our results are central to the general design of liquid-infiltrated dielectric resonators. {\highlight In addition to the efforts put into the cavity structural design (e.g. the optimization of $Q_0$), the present analysis illustrates the importance of choosing the appropriate combination of light-sources ($\lambda$), resonator designs~($f$), materials ($\eta_2$), and bio-chemical buffer media ($\eta_1$).}

\section*{Acknowledgements}

This work is supported by the Danish Council for Strategic Research (DSF Grant No.
2117-05-0037) and the Danish Council for Technology and Production Sciences (FTP Grants No. 274-07-0379 and No. 274-06-0193).

%\bibliography{sensor_references}
%\bibliographystyle{mdpi}
\makeatletter
\renewcommand\@biblabel[1]{#1. }
\makeatother

\newpage

\end{document}